\documentclass[reprint,amsmath,amssymb,aps,]{revtex4-1}
\usepackage{graphicx}
\usepackage{dcolumn}
\usepackage{bm}

\usepackage{amsthm,amsmath,amssymb}
\usepackage{mathrsfs}

\usepackage{floatrow}
\usepackage{makecell,rotating,multirow,diagbox,booktabs}
\usepackage{booktabs}
\usepackage{array, caption, threeparttable}

\begin{document}
\title{ Based-nonequilibrium-environment non-Markovianity, quantum Fisher information and quantum coherence}
\author{Danping Lin}
\author{Hong-Mei Zou}%
\email{zhmzc1997@hunnu.edu.cn}
\author{Jianhe Yang}
\affiliation{Synergetic Innovation Center for Quantum Effects and Application, Key Laboratory of Low-dimensional Quantum Structures and Quantum Control of Ministry of Education, School of Physics and Electronics, Hunan Normal University, Changsha, 410081, P.R. China.}%

\begin{abstract}
In this work, we investigate the non-Markovianity, quantum Fisher information (QFI) and quantum coherence of a qubit in a nonequilibrium environment and have obtained the expressions of QFI and quantum coherence as well as their relationship. We have also discussed in detail the influences of the different noise parameters on these quantum sffects. The results show that the suitable parameters of the nonequilibrium  environment can retard the QFI and quantum coherence in both Markovian and non-Markovian regions. In addition, the smaller memory effects and the larger the jumping rate, the greater the QFI and quantum coherence. And a larger QFI naturally corresponds to a larger quantum coherence, which indicates that the quantum coherence can enlarge the QFI and can effectively enhance the quantum metrology.

\begin{description}
	\item[PACS numbers]
	03.65.Yz, 03.67.Lx, 42.50.-p, 42.50.Pq.
	\item[Keywords]
	Non-Markovianity; Quantum Fisher information; Quantum Coherence; Nonequilibrium environment
\end{description}

\end{abstract}

\maketitle

\section{Introduction}

The parameter estimation is a very basic and important content in quantum information theory. Quantum metrology can attain a measurement precision that surpasses the classically achievable limit by using quantum effects. As we know, the metrology precision can be raised by increasing quantum Fisher information (QFI)\cite{Fisher}. And the Braunstein-Caves theorem proved that the QFI basing on the symmetric logarithmic derivative(SLD) contains all possible information\cite{Braunstein}. On the other hand, quantum coherence(QC) is 

As we all knows, the memory effect of the environment is caused by the long time correlation between the system and the environment and the environment can be divided into Markovian and non-Markovian types according to the memory effects. Many methods for quantifying non-Markovian features have been proposed, such as Rivas-Huelga-Plenio (RHP) measure\cite{Rivas}, Breuer-Laine-Piilo (BLP) measure\cite{Breuer} and Luo-Fu-Song (LFS) measure\cite{Shunlong}. Besides, Zhi He et al.\cite{Zhi} has used the BLP measurement to calculate the non-Markovianity for single-channel open systems, and they obtain a very tight lower bound of non-Markovianity. Yu Liu et al.\cite{Yu} has calculated the non-Markovianity in the dissipative cavity and used it to explain the dynamical behavior of quantum coherence.

Recently, the influence of environments on the QFI and quantum coherence of quantum systems has become an important research topic. ...... one of our authors have obtained the analytical solution of QFI through the SLD method and have enhanced or retarded the QFI by the dissipation cavity\cite{lin}.
Very recently, more and more attention has been paid to study the relationship between different quantum effects. For examples, Zou and Fang have discussed that the discord and entanglement in non-Markovian environments at finite temperatures\cite{zou1}; Zou has discussed the influence of the non-Markovian effect and detuning on the lower bound of the quantum entropic uncertainty relation and entanglement witness in detail\cite{zou2}; Hou has studied the relationship between quantum discord and coherence\cite{Hou}. Feng has verified that the QFI can be utilized to quantify the quantum coherence\cite{Feng}. Wang has shown that quantum coherence also increases the precision of parameter estimation\cite{wang}.

Although many important progresses have been acquired in experimental and theoretical researches on the quantum Fisher information and quantum coherence of open quantum systems, these investigations mentioned above are mainly focused on the Lorentzian and Ohmic environments. Actually, the nonequilibrium environment of dichotomic nature has been observed in some experiments\cite{Taubert,Nowack,Granger}, and the steady-state entanglement and coherence of two coupled qubits\cite{Wang1} and the decoherence induced by non-Markovian noise\cite{X} in a nonequilibrium environment are investigated. In our work, we will study the non-Markovianity, QFI and quantum coherence of a qubit in the nonequilibrium environment. Our main purpose are to understand whether these quantum effects of a qubit in a nonequilibrium environment are different from those in a Lorentzian and an Ohmic environments, and to understand how nonequilibrium environment parameters affect these quantum effects. In addition, we also hope to get the relationship between the QFI and quantum coherence. Hence, we believe that the study of the non-Markovianity, QFI and quantum coherence of a qubit in the nonequilibrium environment is also important and meaningful

The outline of the paper is organized as follows. In Section II, we give the model of a qubit in a nonequilibrium environment. In Section III, we calculate the non-Markovianty, QFI and quantum coherence of the qubit in a nonequilibrium environment. In Section IV, we analyze the influence of the nonequilibrium parameter, the memory effects and the jumping rate on the non-Markovianity, QFI and coherence. Finally, we end with a brief summary of important results in Section V.

\section{Physical Model}

We consider a qubit coupled to a nonequilibrium environment. The Hamiltonian is written as\cite{Xiang}
\begin{equation}\label{EB01}
H=\frac{\hbar}{2}[\omega_{0}+\xi(t)]\sigma_{z},
\end{equation}
where $\sigma_{z}$ and $\omega_{0}$ are the Pauli matrix and the intrinsic transition frequency for the qubit, respectively. $\xi(t)$ denotes the environmental noise caused by environmental effects, which is originally introduced by A. Fuli\'{n}ski\cite{Fulinski}. For simplify, we suppose that $\xi(t)$ is subject to a nonstationary and non-Markovian stochastic process in this paper.

The time evolution of the system is described by the Liouville equation
\begin{equation}\label{EB02}
\frac{\partial}{\partial t}\rho(t;\xi(t))=-\frac{i}{\hbar}[H(t),\rho(t;\xi(t))],
\end{equation}
where the notation $\rho(t;\xi(t))$ indicates the density operator under the environmental noise $\xi(t)$. The reduced density operator of the qubit can be derived by taking an average over the environmental noise as $\rho(t)=\langle\rho(t;\xi(t))\rangle$. By solving Eq.~(\ref{EB02}), the density matrix in the basis $\{|e\rangle,|g\rangle\}$ can be given in the following from
\begin{equation}\label{EB03}
\rho(t)=\left(
\begin{array}{cccc}
\rho_{ee}(0)&\rho_{eg}(0)e^{-i\omega_{0}t}G^{*}(t)\\
\rho_{ge}(0)e^{i\omega_{0}t}G(t)&\rho_{gg}(0)\\

\end{array}
\right),
\end{equation}
where the $G(t)=\langle e^{i\int_{0}^{t}dt^{'}\xi(t)}\rangle$ is a complex time-dependent function.

According to Ref.\cite{X}, the environmental noise $\xi(t)$ is subject to a nonstationary random telegram process that randomly jumps back and forth between $-\nu$ and $+\nu$ at an average rate of $\lambda$. The noise process satisfies the conditional probability such as $\frac{\partial}{\partial t}P(\pm\nu,t|\xi',t')=\mp\int^{t}_{t'}dt 'K(t-\tau)[\lambda P(\nu,\tau|\xi',t')-\lambda P(-\nu|\xi',t')]$, where $K(t -\tau)$ represents a generalized memory kernel. By deriving the calculation, the initial condition of this work is set to $P(\xi_{0})=2^{-1}[(1\pm a)\delta_{\xi_{0,\pm\nu}}]$. And the memory kernel $K$ satisfies $K(t-\tau)=\kappa e^{-\kappa(t-\tau)}$. The environment noise can be characterized by the nonequilibrium parameter $a$ ($\mid a\mid\leq1$) and the memory decay rate $\kappa (\kappa>0)$ as well as the amplitude switches randomly with the jumping rate $\lambda$ between the values $\pm\nu(\nu>0)$. $a=0$ indicates that the environment is equilibrium (and $a\neq0$ vice versa). The smaller $\kappa$ and bigger $\nu$ represent the stronger non-Markovianity. Under nonequilibrium environment, $G(t)$ in Eq.~(\ref{EB03})can be analytically solved as\cite{X}
\begin{equation}\label{EB04}
G(t)=\mathcal{L}^{-}[G(s)],\\
G(s)=\frac{s^{2}+(\kappa+ia\nu)s+\kappa(2\lambda+ia\nu)}{s^{3}+\kappa s^{2}+(2\kappa\lambda+\nu^{2})s+\kappa\nu^{2}},
\end{equation}
where $\mathcal{L}^{-}$ indicates the inverse Laplace transform and the initial condition is $G(0)=1$.

Using $G(0)=1$, one can obtain easily $G(s)=\frac{(s-u_{+})(s-u_{-})}{(s-s_{1})(s-s_{2})(s-s_{3})}$ where $u_{\pm}$ are roots of $s^{2}+(\kappa+ia\nu)s+\kappa(2\lambda+ia\nu)=0$, and $s_{j}(j=1,2,3)$ are roots of $s^{3}+\kappa s^{2}+(2\kappa\lambda+\nu^{2})s+\kappa\nu^{2}=0$

We can acquire the exact solution of $G(t)$ as\cite{Piessens}, i.e.
\begin{equation}\label{EB08}
G(t)=\sum_{j=1}^{3}G(s)e^{s_{j}t}.
\end{equation}

Let an arbitrary initial state is $|\psi\rangle=\cos\frac{\theta}{2}|e\rangle+e^{i\phi} \sin\frac{\theta}{2}|g\rangle$. The reduced
density matrix $\rho(t)$ in Eq.~(\ref{EB03}) can write as
\begin{equation}\label{EB11}
\rho(t)=\left(
\begin{array}{cccc}
\rho_{ee}(t)&\rho_{eg}(t)\\
\rho_{ge}(t)&\rho_{gg}(t)\\

\end{array}
\right),
\end{equation}
with the elements

\begin{equation}\label{EB12}
\begin{split}
\rho_{ee}(t)=&\frac{1}{2}(1+\cos\theta),\\
\rho_{eg}(t)=&\frac{1}{2}\sin\theta e^{-i(\phi+\omega_{0}t)}G^{*}(t),\\
\rho_{ge}(t)=& \rho^{\ast}_{eg}(t),\\
\rho_{gg}(t)=&1- \rho_{ee}(t).
\end{split}
\end{equation}

\section{Non-Markovianity, QFI and Quantum Coherence}

In the three subsections, we briefly review the measurements of the non-Markovianity, QFI and quantum coherence.

\subsection{ non-Markovianity}
\label{sec:3}
In this part, we introduce the non-Markovianity of the qubit by using the BLP measure, which based on the trace distance between two quantum states $\rho_{1}$ and $\rho_{2}$\cite{Breuer}. The non-Markovianity $\mathcal{N}$ is defined as
\begin{eqnarray}\label{EB22}
\begin{array}{cccc}
\mathcal{N} &=& \max\limits_{\rho_{1}(0),\rho_{2}(0)}\int_{\sigma>0}\sigma[t,\rho_{1}(0),\rho_{2}(0)]dt
\end{array},
\end{eqnarray}
where $\sigma[t,\rho_{1}(0),\rho_{2}(0)]$ is the change rate of the trace distance defined as $\sigma[t,\rho_{1}(0),\rho_{2}(0)]=\frac{d}{dt}D[\rho_{1}(t),\rho_{2}(t)]$, and the trace distance is defined as $D[\rho_{1}(t),\rho_{2}(t)]=\frac{1}{2}Tr|\rho_{1}(t)-\rho_{2}(t)|$ with $|A|=\sqrt{A^{\dag}A}$ and $0\leq D\leq1$. If $D=0$, the two state are the same, and if $D=1$, the two state are totaly distinguishable. That is, when $\rho_{1}(0)=|e\rangle\langle e|$ and $\rho_{2}(0)=|g\rangle\langle g|$, the trace distance is maximum.

\subsection{Quantum Fisher Information}

Based on thee classical Fisher information (CFI)\cite{C,A,Pairs}, the quantum Cram\'{e}r-Rao (QCR) theorem of the QFI has also be proposed, which is $\langle(\bigtriangleup \hat{\lambda})^{2}\rangle_{\lambda}\geq\frac{1}{N \mathcal{F}_{\lambda}}$, where $\hat{\lambda}$ is the unbiased estimator of parameter $\lambda$. The QFI theorem gives the highest accuracy of parameter estimation in the case of quantum mechanics. The smaller the $\langle(\bigtriangleup \hat{\lambda})^{2}\rangle_{\lambda}$, the larger the $\mathcal{F}_{\lambda}$. That is, we can get the higher the quantum metrology precision.

Using positive-operator valued measurement (POVM), $\mathcal{ F}_{\lambda}$ can be written as
\begin{eqnarray}\label{EB15}
\mathcal{ F}_{\lambda}&=&Tr(\rho_{\lambda}L_{\lambda}^2)=Tr(\partial_{\lambda}\rho_{\lambda}L_{\lambda}),
\end{eqnarray}
this is quantum Fisher information (QFI), where $L_{\lambda}$ is symmetric logarithmic derivatives(SLD) for the parameter $\lambda$, which is a Hermitian operator determined by $\partial_{\lambda}\rho_{\lambda}=\frac{1}{2}\{\rho_{\lambda},L_{\lambda}\}$ where $\partial_{\lambda}\equiv\frac{\partial}{\partial\lambda}$ and $ \{\cdot,\cdot\}$ denotes the anticommutator. If the density operator $\rho_{\lambda}$ satisfies the spectral decomposition $\rho_{\lambda}=\sum_{i}p_{i}|\psi_{i}\rangle\langle\psi_{i}|$ and without considering the existence of zero eigenvalues, the $\mathcal{F}_{\lambda}$ can be expressed as follows
\begin{eqnarray}\label{EB17}
\begin{array}{cccc}
\mathcal{F}_{\lambda}&=&\sum_{i^{'}}\frac{(\partial_{\lambda}p_{i^{'}})^{2}}{p_{i^{'}}}+2\sum_{i\neq j}\frac{(p_{i}-p_{j})^{2}}{p_{i}+p_{j}}|\langle\psi_{i}|\partial_{\lambda}\psi_{j}\rangle|^{2}\\
\end{array},
\end{eqnarray}

Using Eq.~(\ref{EB15}) to calculate the QFIs of parameter $\theta$ and $\phi$. Inserting  Eq.~(\ref{EB11}) into Eq.~(\ref{EB15}), the QFIs are obtained as
\begin{eqnarray}\label{EB19}
\begin{array}{cccc}
\mathcal{F_{\theta}}&=&1\\
\mathcal{F_{\phi}}&=&|G(t)|^{2}\sin^{2}\theta\\
\end{array},
\end{eqnarray}
From Eq.~(\ref{EB19}), it is known that $\mathcal{F_{\theta}}$ is always 1 and the $\mathcal{F_{\phi}}$ is determined by two factors, $G(t)$ and $\theta$.

\subsection{ Quantum Coherence}

A reasonable measure to quantify quantum coherence should fulfill the following condition: non-egativity, monotonicity, strong monotonicity, convexity, uniqueness for pure states and additivity\cite{Streltsov}. According to a set of properties of quantum coherence measure, some coherence measures are proposed. Here, we focus on the $l_{1}$ norm of coherence\cite{Baumgratz}. From Eq.~(\ref{EB11}) it can be expressed as
\begin{eqnarray}\label{EB20}
\mathcal{C}_{l}(t) &=& \sum_{i,j(i\neq j)}|\rho_{ij}(t)| = |G(t)||\sin\theta|,
\end{eqnarray}
where $\mid\rho_{ij}(t)\mid (i\neq j)$ is the absolute value of the non-diagonal elements $\rho_{ij}(t)$ of the density matrix $\rho(t)$.

Comparing Eq.~(\ref{EB19}) and Eq.~(\ref{EB20}), we can obtain the analytical relationship between the $\mathcal{F_{\phi}}$ and the $\mathcal{C}_{l}(t)$, which is
\begin{eqnarray}\label{EB21}
\mathcal{F_{\phi}}&=& \mathcal{C}^{2}_{l}(t),
\end{eqnarray}

From Eq.~(\ref{EB20}), we know that the QFI will enlarge with the quantum coherence increasing, which indicates that the quantum coherence can augment the QFI and can effectively enhance the quantum metrology.

\section{Discussion and results}

In this section, we illustrate the influence of environmental parameters (the nonequilibrium parameter $a$, the memory decay rate $\kappa$ and the jumping rate $\lambda$) on non-Markovianity, QFI and quantum coherence. The quantum system exhibits a Markovian behavior when $(\nu/\lambda<1)$, the evolution of the quantum system is non-Markovian only when $(\nu/\lambda>1)$\cite{Xiang}.

In Fig. 1, we plot the non-Markovianity $\mathcal{N}$ as function of the dimensionless quantity $\nu/\lambda$, which shows that $\mathcal{N}$ is dependent on $\nu/\lambda$. However, due to the difference between the selected parameters and the calculation method, the curve change of $\mathcal{N}$ is different from that of the Ref.\cite{Xiang}. In short, $\mathcal{N}$ increases as $\nu/\lambda$ increases. For the same equilibrium parameter $a$, the larger the value of $\nu$ is, the larger $\mathcal{N}$ is. For the same the value of $\nu/\lambda$ , the smaller the equilibrium parameter $a$, the larger $\mathcal{N}$. As is plotted in Fig. 2, $\mathcal{N}$ associated to the memory decay rate $\kappa/\lambda$. $\mathcal{N}$ decreases as $\kappa/\lambda$ increases, and $\mathcal{N}$ will eventually disappear when $\kappa\rightarrow\infty$. When $\nu=0.8\lambda$ and $\kappa/\lambda=0$, the maximum value of $\mathcal{N}$ can reach 4.0, indicating that the non-Markovian effect is very obvious. When $\nu=0.8\lambda$ and $\kappa/\lambda=10$, the non-Markovianity is zero. The dynamic behavior of the qubit presents Markovian characteristics. For the same $\nu$, the smaller $\kappa$, the larger $\mathcal{N}$. And, the non-Markovianity has the similar dynamic behaviors in the cases of $\nu=0.8\lambda$ and $\nu=4\lambda$, while the $\mathcal{N}$ is larger in the former than in the latter.

\begin{figure}[htbp]
	\includegraphics[width=5.5cm]{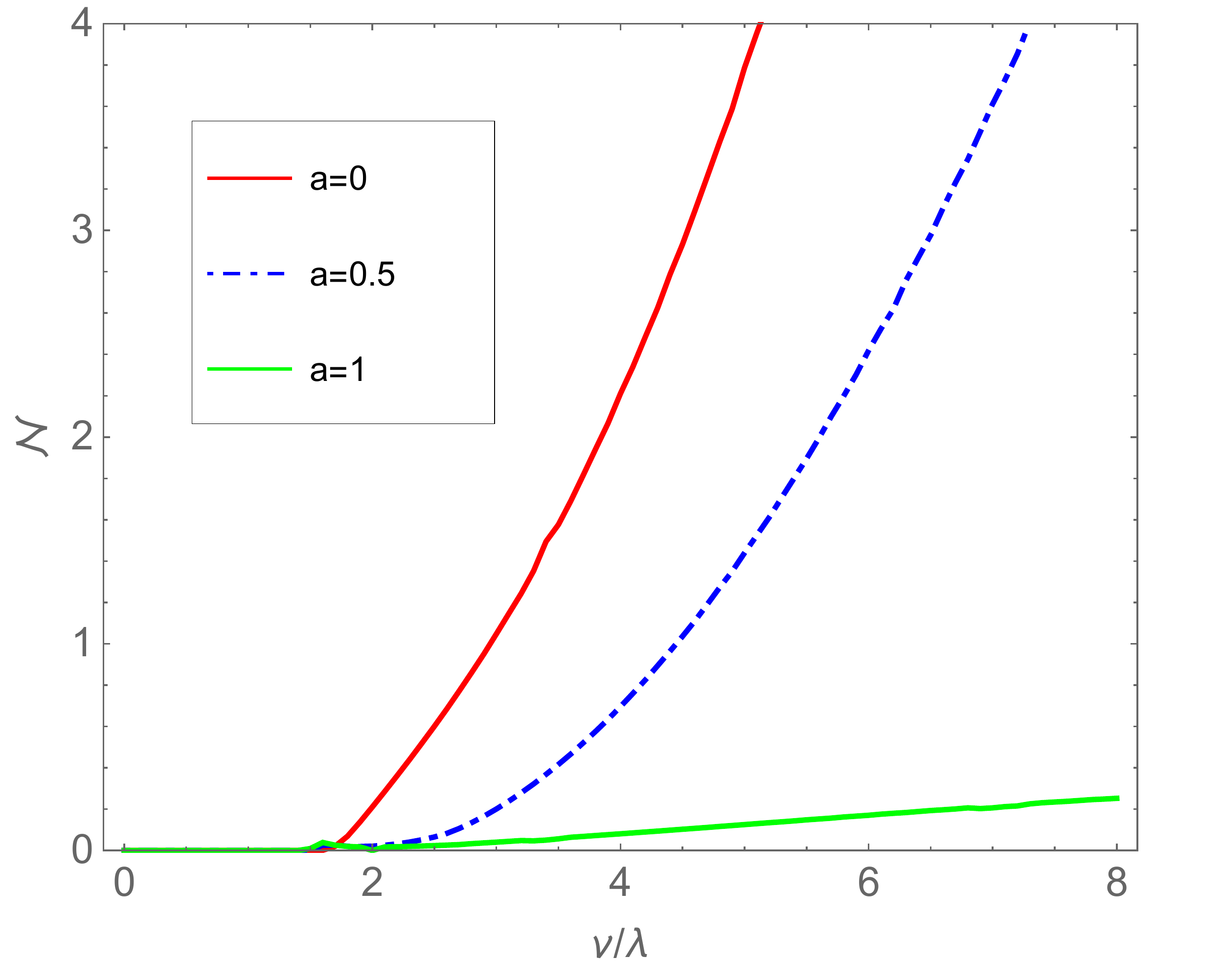}
	\caption{(Color online) Non-Markovianity $\mathcal{N}$ as a function of the dimensionless quantity $\nu/\lambda$. Here $\theta=\frac{\pi}{2},\phi=0$. The memory decay rate $\kappa=8\lambda.$}
	\label{fig:1}
\end{figure}

\begin{figure}[htbp]
	\includegraphics[width=5.5cm]{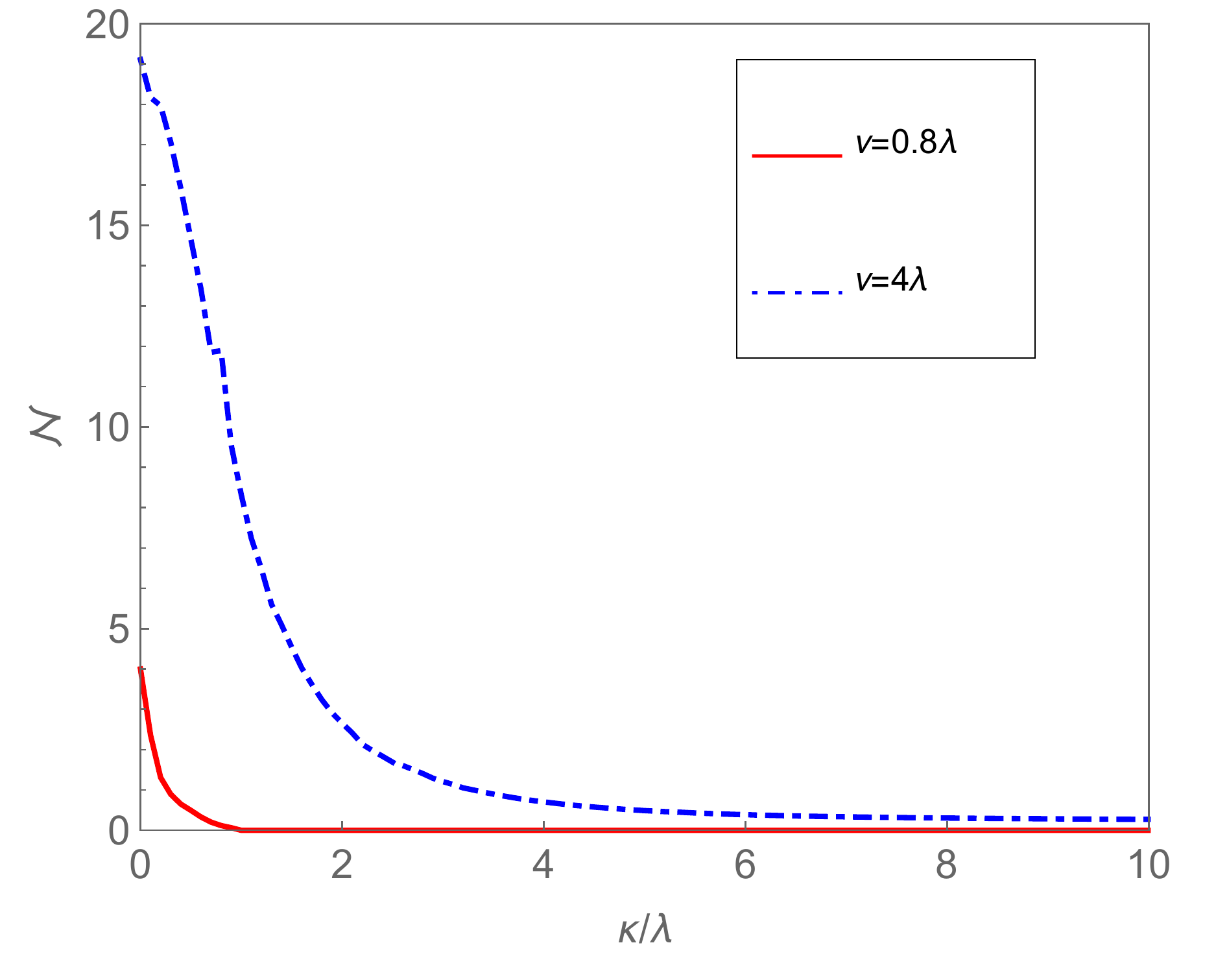}
	\caption{(Color online) Non-Markovianity $\mathcal{N}$ as a function of the dimensionless quantity $\kappa/\lambda$. Here $\theta=\frac{\pi}{2}, \phi=0$, $a=0.5$.}
	\label{fig:2}
\end{figure}

\begin{figure}[htbp]
	\includegraphics[width=5.5cm]{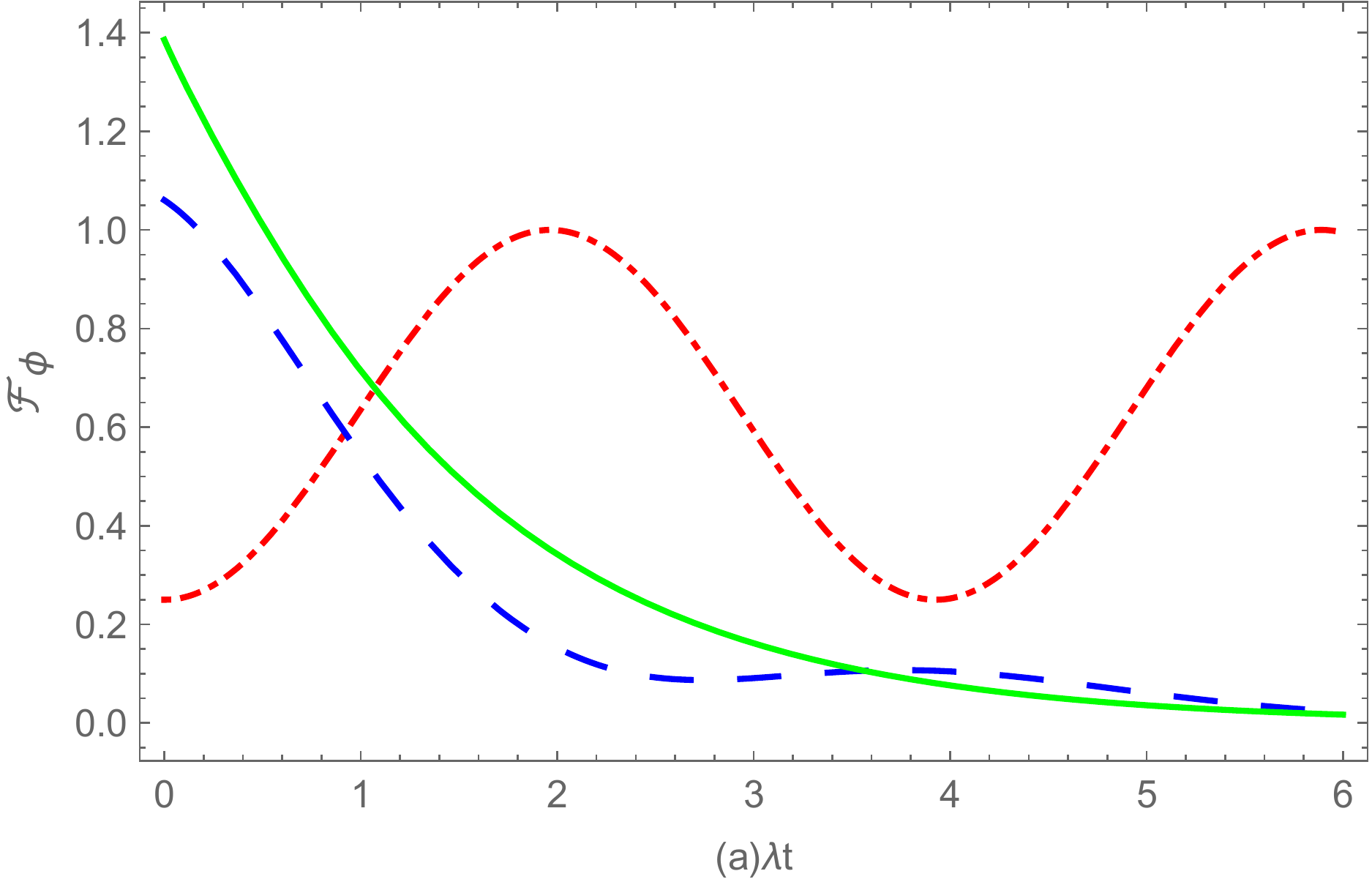}
	\includegraphics[width=5.5cm]{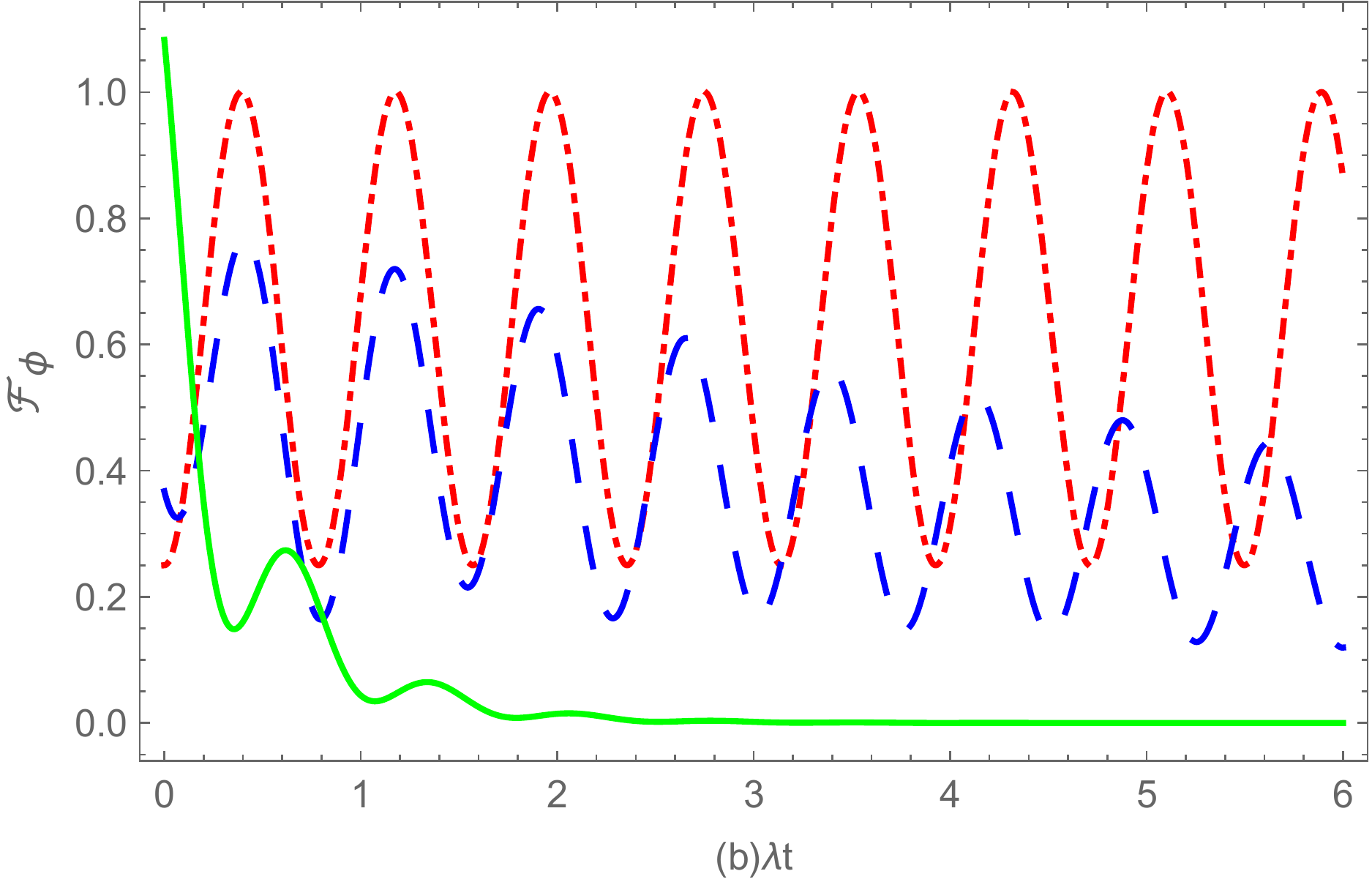}
	
	\includegraphics[width=5.5cm]{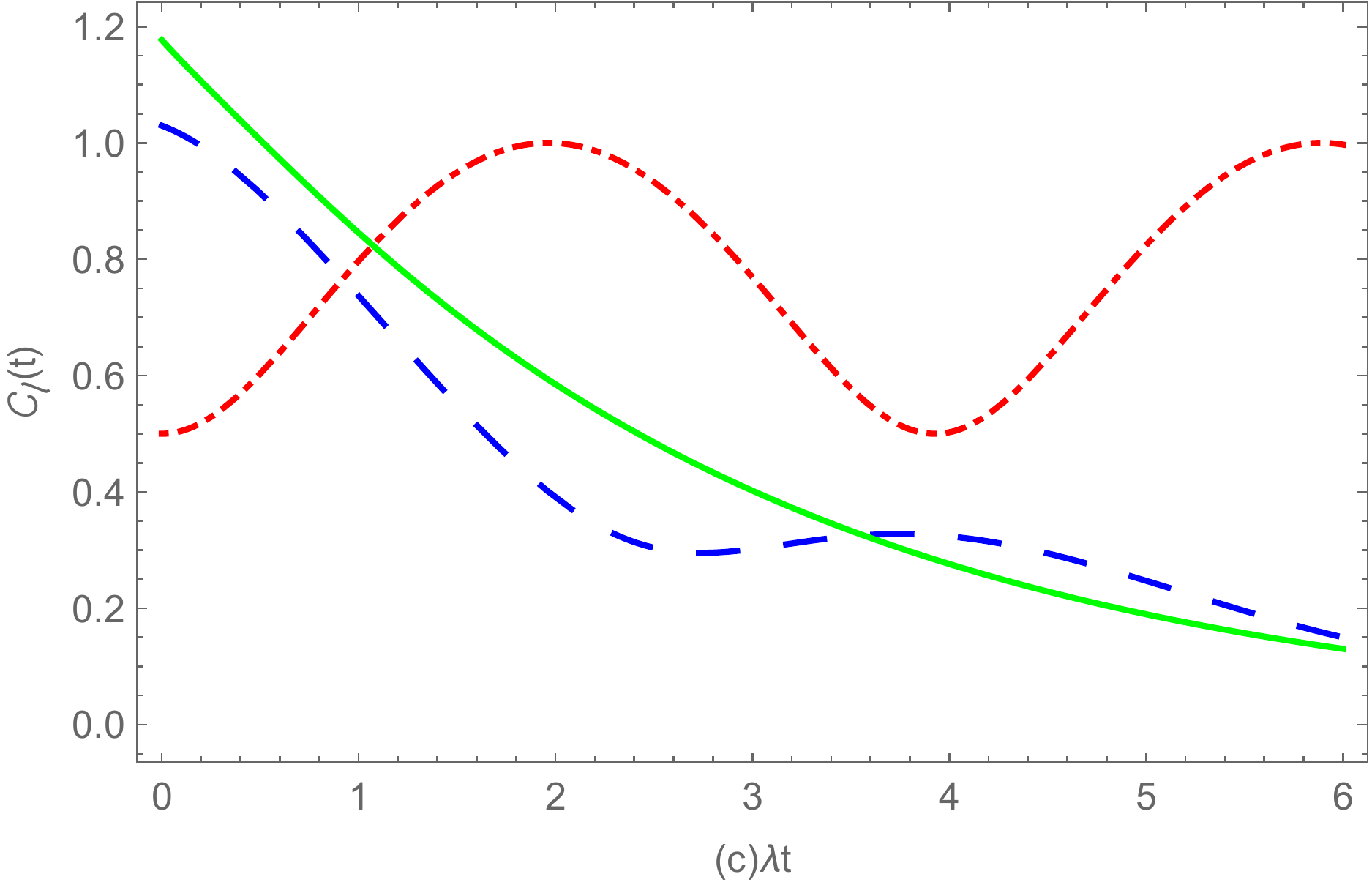}
	\includegraphics[width=5.5cm]{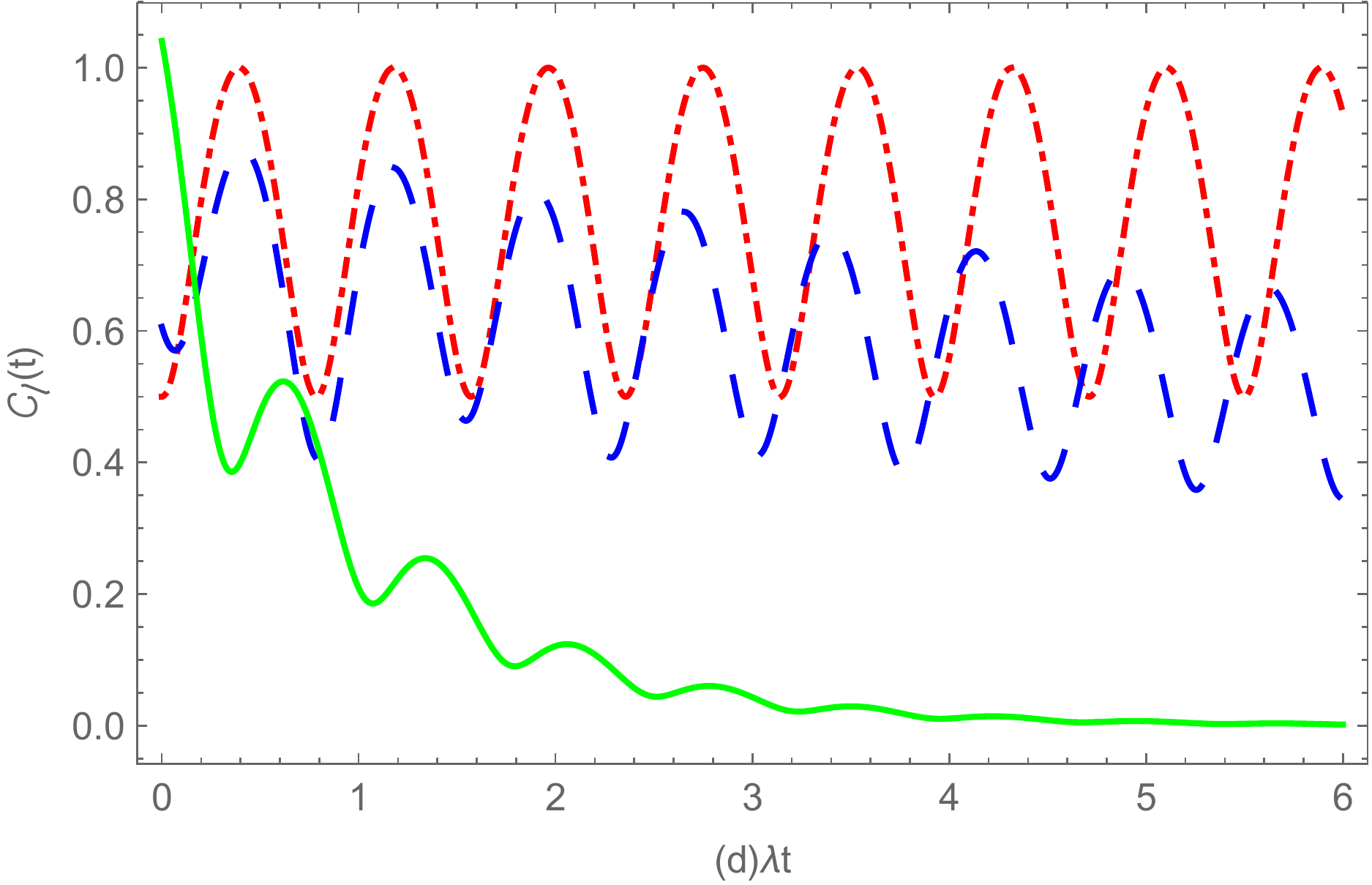}
	\caption{(Color online) QFI $\mathcal{F}_\phi$ and quantum coherence $\mathcal{C}_{l}$ as a function of $\lambda t$ for $a=0.5$ and different values of the memory decay rate $\kappa$: $\kappa=0$ (red dot dash line), $\kappa=\lambda$ (blue dotted line), $\kappa=10\lambda$ (green line). (a) and (c)$\nu=0.8\lambda$; (b) and (d)$\nu=4\lambda$. The other parameter are $\theta=\frac{\pi}{2}$ and $\phi=0$.}
	\label{fig:3}       
\end{figure}

Fig. 3 is plotted $\mathcal{F}_\phi$ and $\mathcal{C}_{l}$ as a function of $\lambda t$ under different values of $\kappa$ and $\nu$ for $a=0.5$. Fig. 3 (a) shows that the $\mathcal{F}_\phi$ has the periodic oscillation when $\kappa=0$ and $\nu=0.8\lambda$. However, when $\kappa=10\lambda$ and $\nu=0.8\lambda$, the $\mathcal{F}_\phi$ decays monotonically, which implies that the $\mathcal{F}_\phi$ dynamics of the qubit becomes Markovian. Therefore, with $\kappa$ decreasing, the evolution of the qubit will change from Markovian to non-Markovian, corresponding to the tendency of $\mathcal{F}_\phi$ from monotonical decay to oscillation. As shown in Fig. 3 (b),  when $\kappa=10\lambda$ and $\nu=4\lambda$, the $\mathcal{F}_\phi$ could also take the behavior of oscillations. In addition, the maximum value of oscillation recovery of $\mathcal{F}_\phi$ increases with the decreasing of $\kappa$, which implies that the smaller the value of $\kappa/\lambda$, the more obvious the non-Markovian effects. In Fig. 3 (c) and (d), $\mathcal{C}_{l}$ is plotted as a function of $\lambda t$ for $a=0.5$ and different values of $\kappa$ and $\nu$. Fig. 3 (c) displays an oscillatory behavior of $\mathcal{C}_{l}$ when $\kappa$ is very small and $\nu=0.8\lambda$. However, $\mathcal{C}_{l}$ decays monotonically for larger $\kappa$, which shows that the dynamical evolution becomes Markovian. Similar to the case of Fig. 3 (c), as $\kappa$ increase, the $\mathcal{C}_{l}$ rapidly oscillations to zero when $\nu=4\lambda$, as shown in the Fig. 3 (d).

\begin{figure}
	\includegraphics[width=5.5cm]{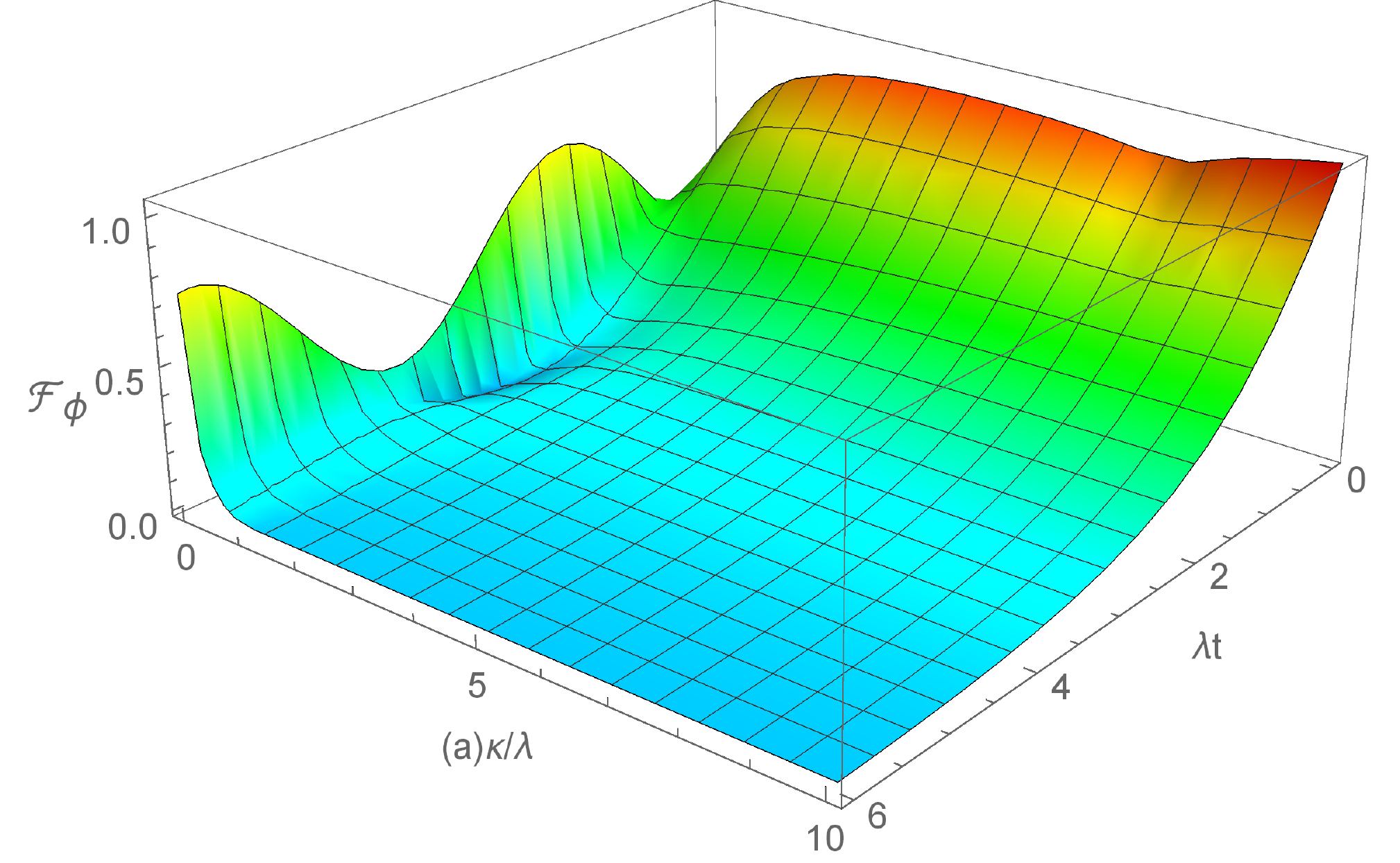}
	\includegraphics[width=5.5cm]{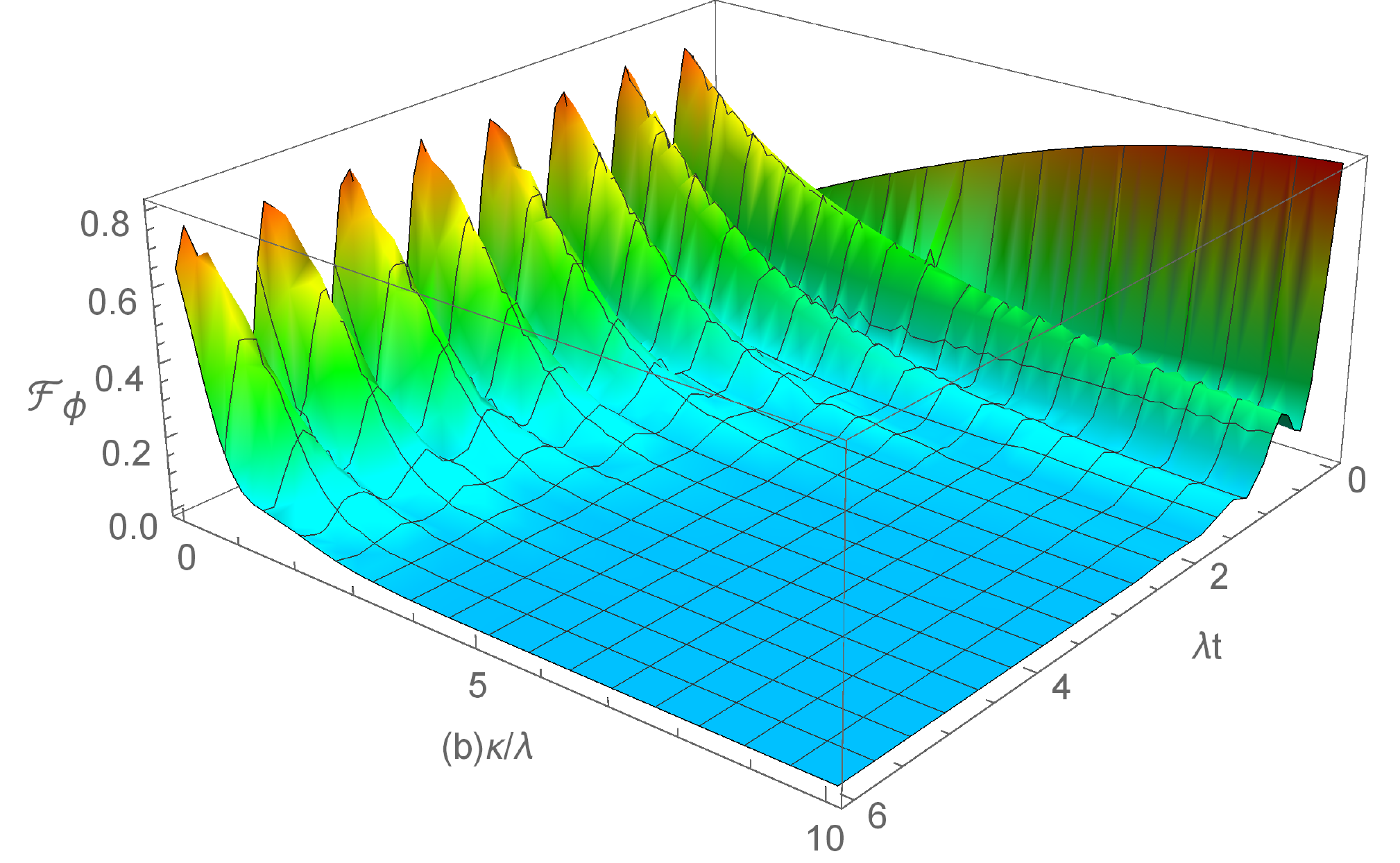}
	\caption{(Color online) QFI $\mathcal{F}_\phi$ as a function of the dimensionless quantity $\lambda t$ and the memory decay rate $\kappa$ for $a=0.5$. Here $\theta=\frac{\pi}{2}$ and $\phi=0$. (a)$\nu=0.8\lambda$; (b)$\nu=4\lambda$.}
	\label{fig:4}       
\end{figure}

In order to observe comprehensively the dependence of $\mathcal{F}_\phi$ on $\kappa$ and $\lambda t$, we can plot a three-dimensional diagram. Fig. 4 is plotted $\mathcal{F}_\phi$ as a function of the dimensionless quantity $\lambda t$ and the memory decay rate $\kappa$ for $a=0.5$. Fig. 4(a) shows that $\mathcal{F}_\phi$ appears the phenomenon of rapidly revivals when $\kappa/\lambda=10\lambda$ and $\nu=0.8\lambda$. However, when $\kappa/\lambda\gg1$ and $\nu=0.8\lambda$, $\mathcal{F}_\phi$ appears decay phenomenon. Fig. 4(b) shows that no matter what $\kappa$ is, $\mathcal{F}_\phi$ appears the phenomenon of oscillations. The memory effect is the reason for the revival of $\mathcal{F}_\phi$. Namely, the memory effect can enhance the non-Markovian behavior and the value of QFI. But, the oscillatory behavior of the $\mathcal{F}_\phi$ does not disappear in the memoryless limit. The three-dimensional picture of $\mathcal{C}_{l}(t)$ is similar to the $\mathcal{F}_\phi$, we omit it in order to reduce the space.

\begin{figure}
	\includegraphics[width=5.5cm]{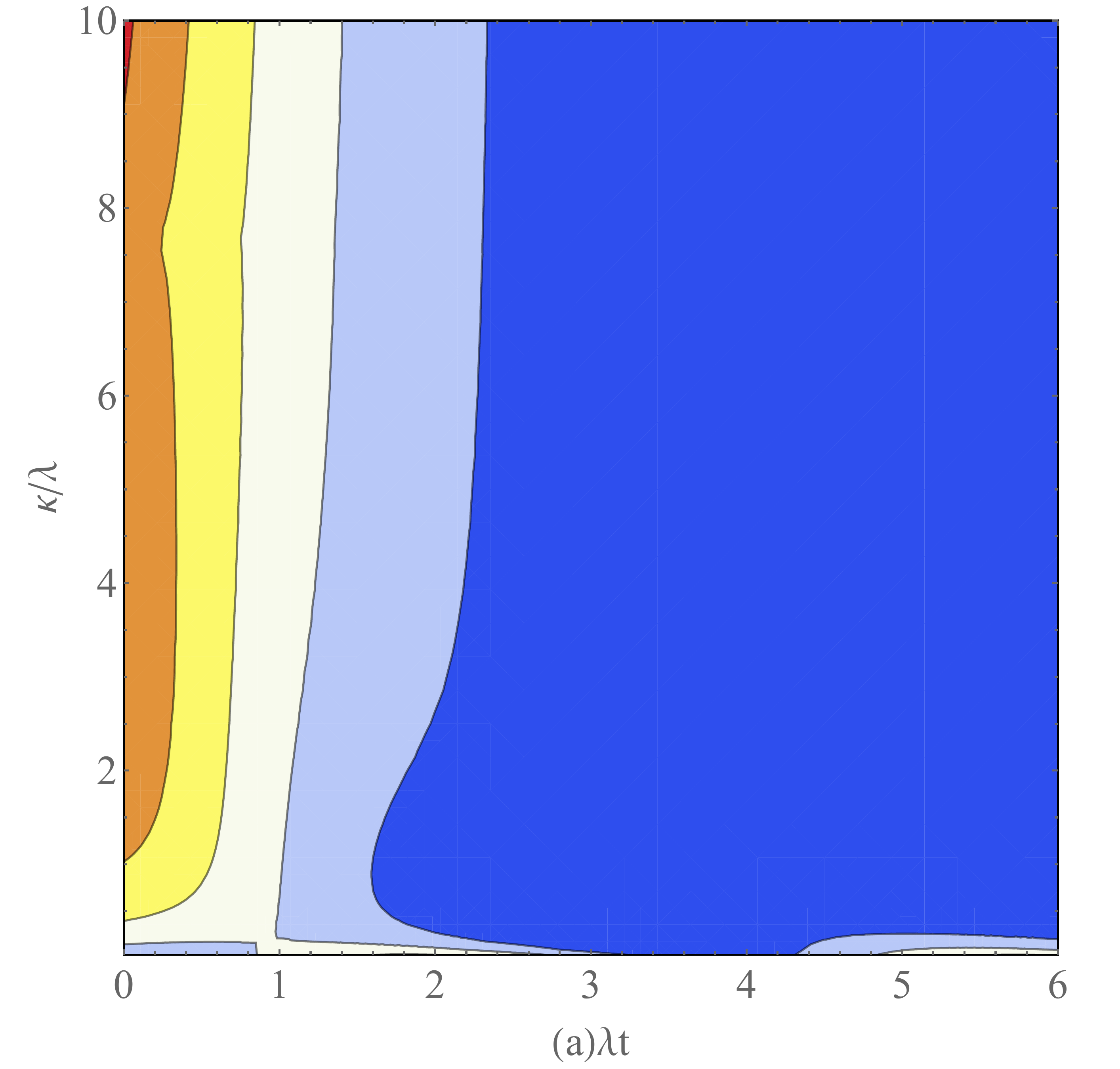}
	\includegraphics[width=5.5cm]{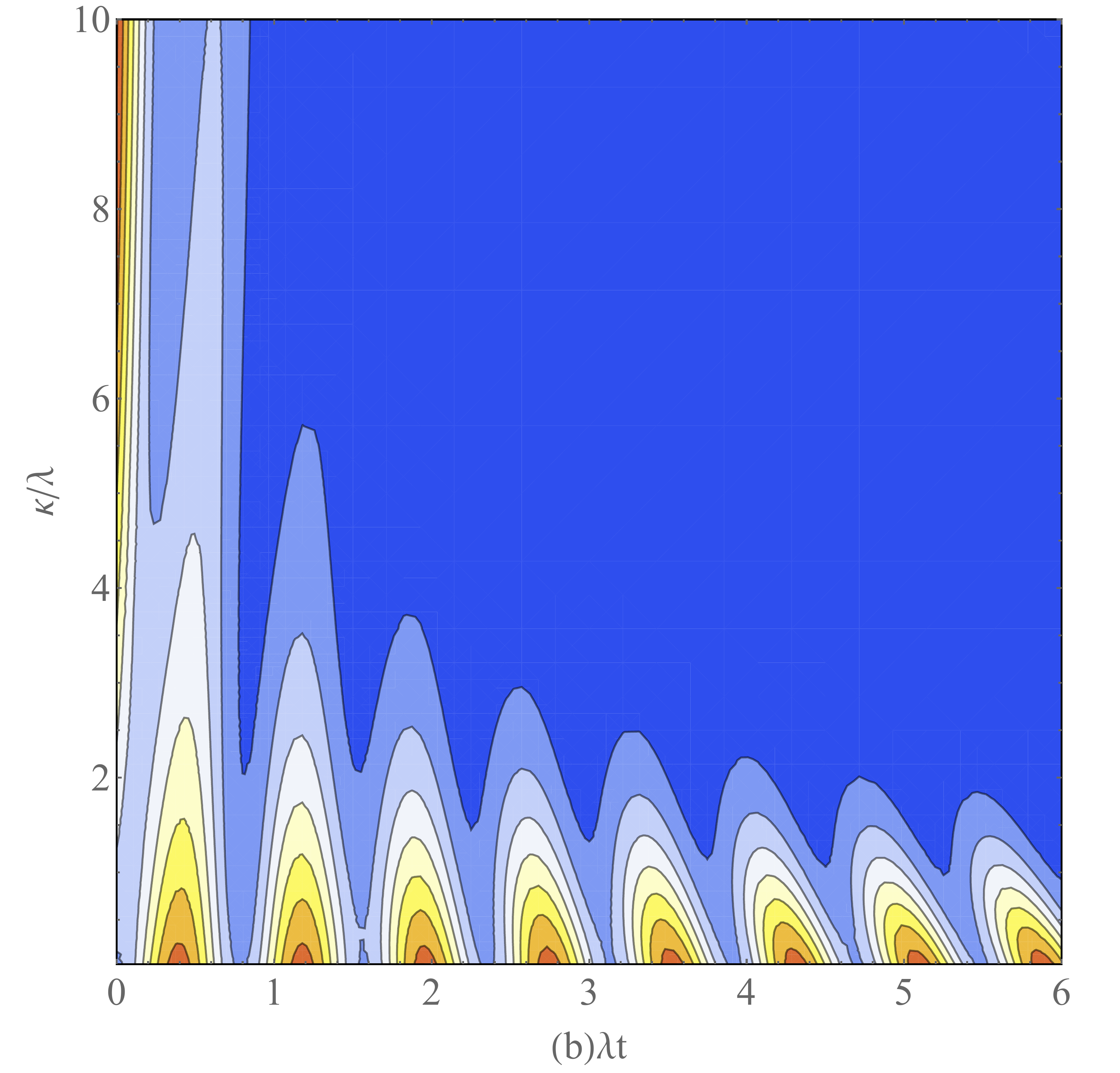}
	\caption{(Color online) $\mathcal{F}_\phi$ as a functions of the dimensionless quantity $\lambda t$ and the memory decay rate $\kappa$ for $a=0.5$. Here $\theta=\frac{\pi}{2}$ and $\phi=0$. (a)$\nu=0.8\lambda$; (b)$\nu=4\lambda$.}
	\label{fig:5}       
\end{figure}

In Fig.5, we plot the $\mathcal{F}_\phi$ as a functions of the dimensionless quantity $\lambda t$ and the memory decay rate $\kappa/\lambda$ for $a=0.5$. From this figure, it can be seen that the smaller the value of $\kappa/\lambda$ and $\lambda t$, the larger the value of $\mathcal{F}_\phi$. Meanwhile, the evolution behavior of $\mathcal{C}_{l}(t)$ is similar to the $\mathcal{F}_\phi$, we omit it in order to reduce the space. Furthermore, as is shown above, the evolution behavior of $\mathcal{C}_{l}$ is similar qualitatively to the $\mathcal{F}_\phi$ in both Markovian and non-Markovian regions. That is, the smaller $\kappa$ is beneficial to the $\mathcal{F}_\phi$ and $\mathcal{C}_{l}$. Comparing Fig. 3 (a) and (c) (or Fig. 3 (b) and (d)), we can see the tendency of $\mathcal{F}_\phi$ and $\mathcal{C}_{l}$ with $\nu=0.8\lambda$ and $\nu=4\lambda$ is coincident.

\begin{figure}
	\includegraphics[width=5.5cm]{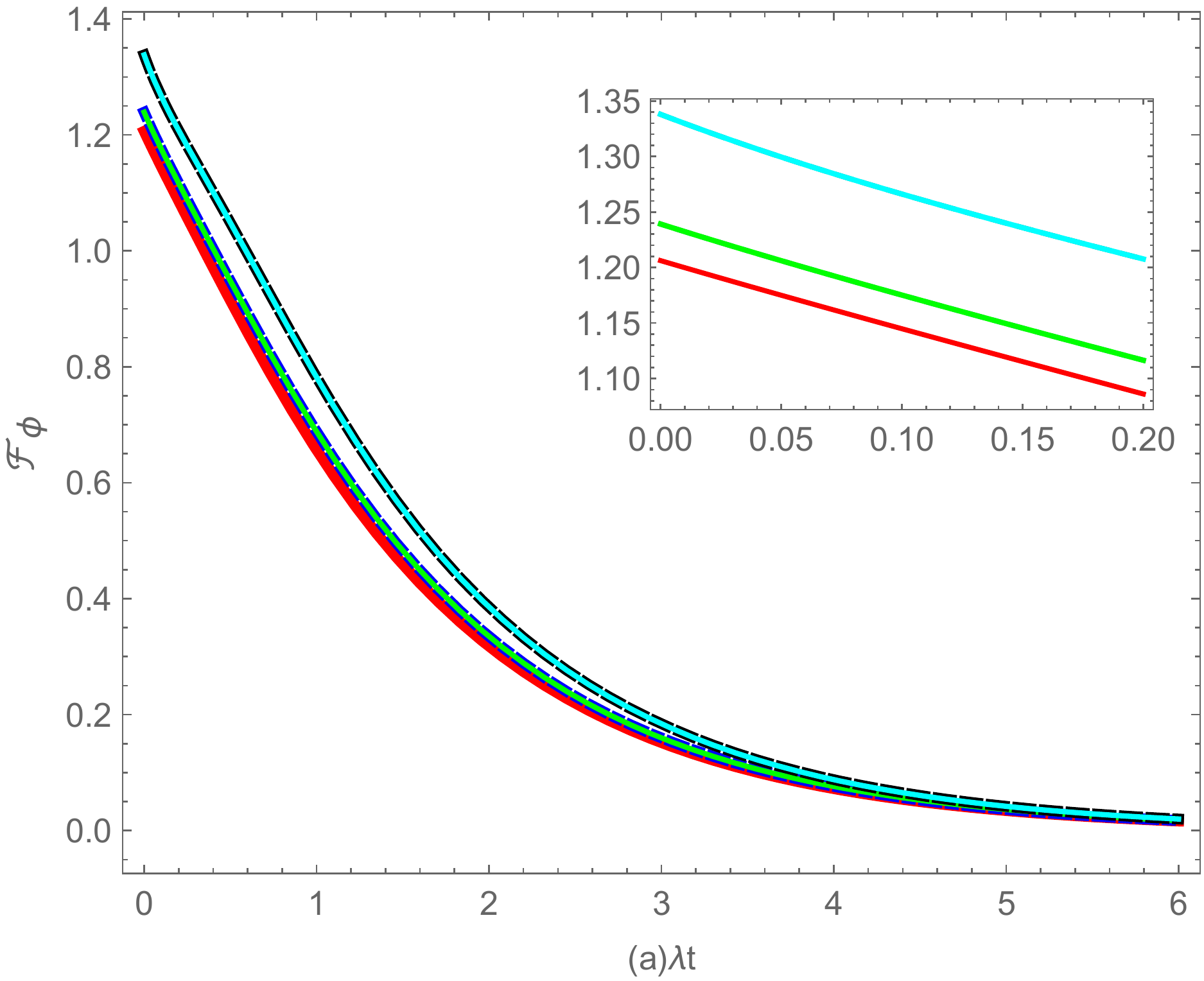}
	\includegraphics[width=5.5cm]{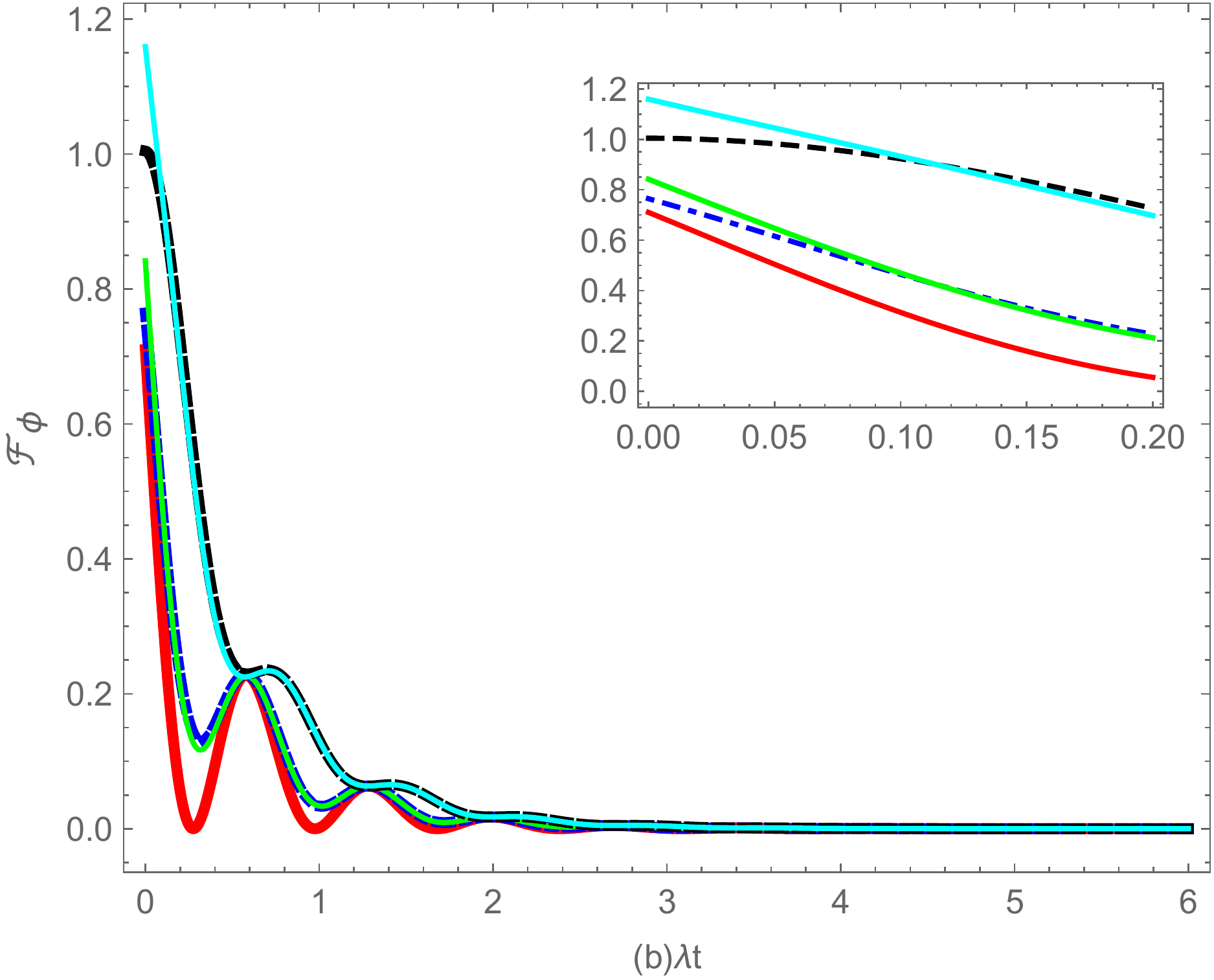}
	
	\includegraphics[width=5.5cm]{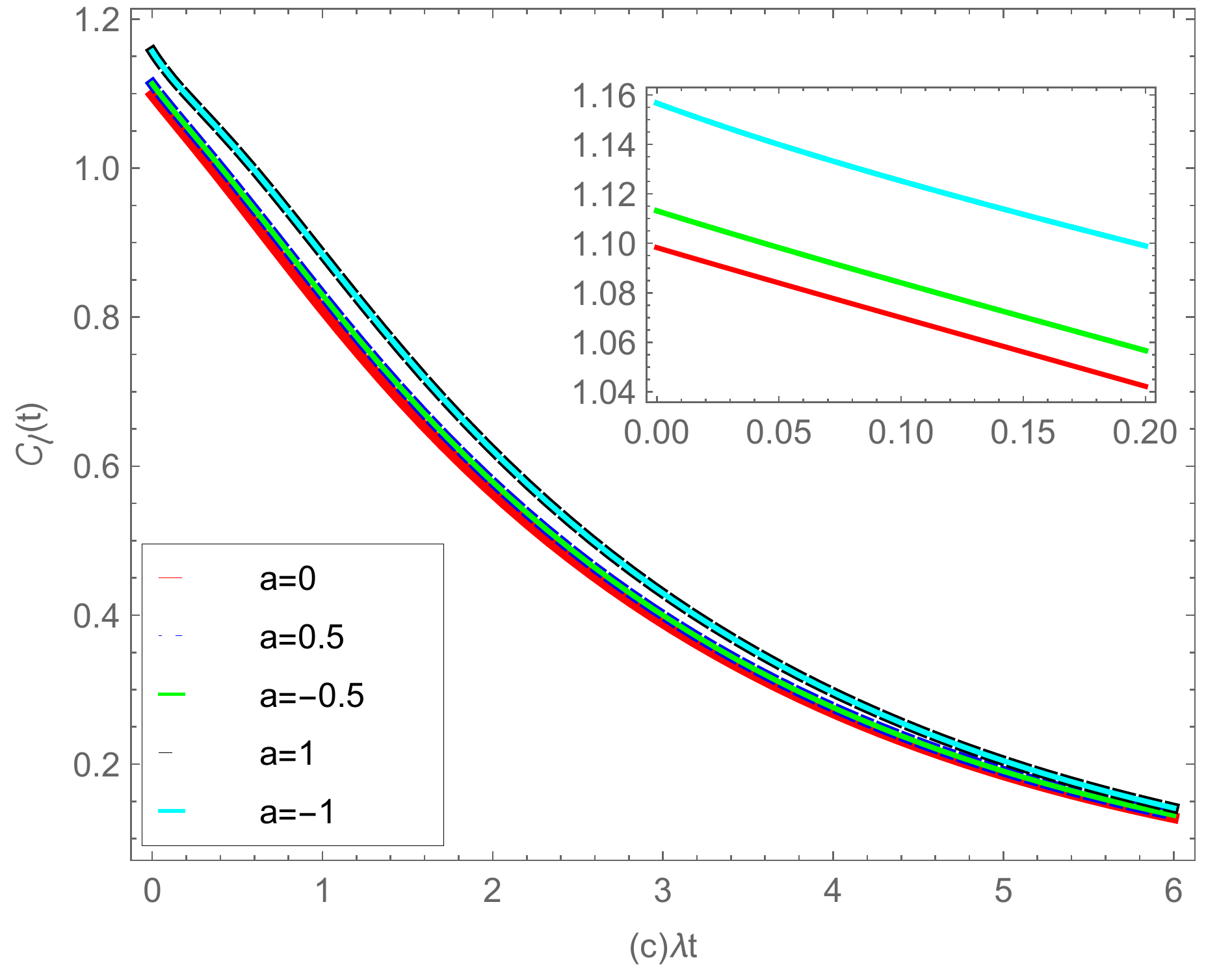}
	\includegraphics[width=5.5cm]{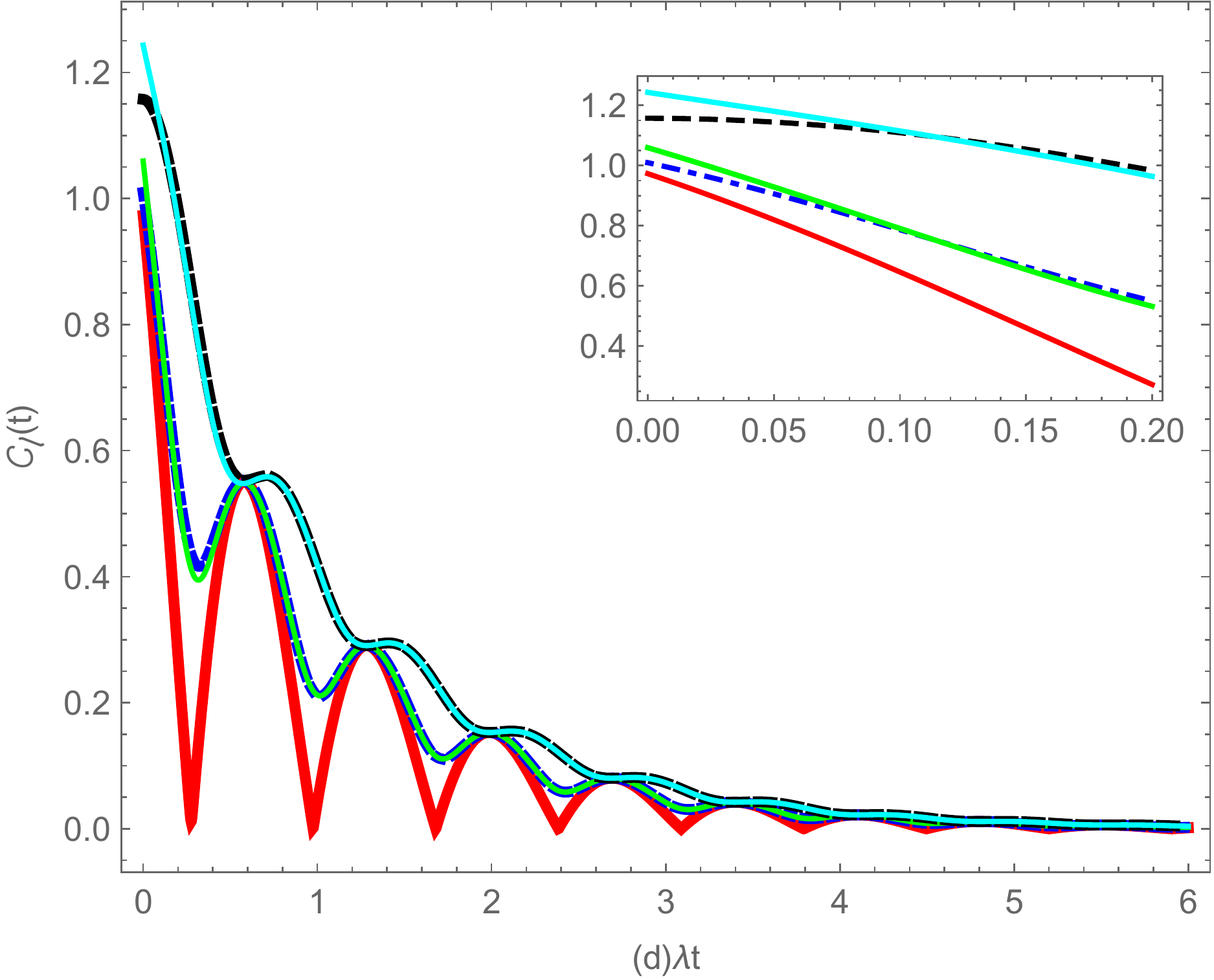}
	\caption{(Color online) QFI $\mathcal{F}_\phi$ and coherence $\mathcal{C}_{l}$ as a function of dimensionless quantity $\lambda t$ for the initial state $|\psi\rangle$. (a), (c)$\nu=0.8\lambda$; (b), (d)$\nu=4\lambda$. The memory decay rate $\kappa=8\lambda.$ The insets show the evolution of a short time. The other parameter are $\theta=\frac{\pi}{2}$ and $\phi=0$.}
	\label{fig:6}       
\end{figure}

In Fig. 6(a), $\mathcal{F}_\phi$ is plotted as a function of the dimensionless quantity $\lambda t$ for the different nonequilibrium parameter $a$ in the Markovian dynamic region ($\nu=0.8\lambda$). As is shown above, there is a little difference among the five lines and $\mathcal{F}_\phi$ decreases monotonically with time $t$ and vanishes in the asymptotic limit $t\rightarrow\infty$, which implies the accuracy of the parameter estimation is smaller. If the environment is out of equilibrium ($0<|a|\leq1$), the greater the value of $\mid a\mid$, the greater the initial value of $\mathcal{F}_\phi$. But they all close to zero when time $t$ increases. And the curves of $\mathcal{F}_\phi$ are coincident for $a=0.5$ and $a=-0.5$, the curves of $\mathcal{F}_\phi$ are also coincident for $a=1$ and $a=-1$. The inset of Fig. 6(a) shows that the initial value of $\mathcal{F}_\phi$ will be larger when the system is far away from the equilibrium environment. In Fig. 6(b), $\mathcal{F}_\phi$ is plotted as a function of the dimensionless quantity $\lambda t$ for the different nonequilibrium parameter $a$ in the non-Markovian dynamic region ($\nu=4\lambda$). It is seen that the initial value of $\mathcal{F}_\phi$ has the different value of the nonequilibrium parameter $a$. $\mathcal{F}_\phi$ appears oscillating recovery phenomenon with time $t$ due to the feedback effects of the non-Makovian environment. Similar to the case of the inset of Fig. 6(a), the initial value of $\mathcal{F}_\phi$ will be larger as $|a|$ increases, as shown the inset of Fig. 6(b).

In Fig. 6(c), $\mathcal{C}_{l}$ is plotted as a function of the dimensionless quantity $\lambda t$ for different the $a$ in the Markovian region ($\nu=0.8\lambda$). In Fig. 7 (d), we plot the time evolution of the $\mathcal{C}_{l}$ for $\theta=\frac{\pi}{2}$, $\phi=0$  and $\nu=4\lambda$. It can be seen that the $\mathcal{C}_{l}$ oscillates damply to zero. This figure shows that the $\mathcal{C}_{l}$ in the nonequilibrium environment is greater than the equilibrium environment in the short time. But the $\mathcal{C}_{l}$ will be close to zero in a long evolution time. Obviously, the nonequilibrium parameter $a$ plays the important role in the dynamic revolutions of the $\mathcal{F}_\phi$ and $\mathcal{C}_{l}$. When $|a|$ is greater, the initial value of coherence is larger, as shown the inset of Fig. 6(d).

As a comparison, when the environment is out of equilibrium, the $\mathcal{C}_{l}$ and $\mathcal{F}_\phi$ will be delay in both Markovian and non-Markovian regions, which indicates that the bigger $\nu$, the slower down the evolution of $\mathcal{C}_{l}$ and $\mathcal{F}_\phi$. For example, comparing Fig. 6(a) and Fig. 6(c), it can be seen that the tendency of $\mathcal{C}_{l}$ and $\mathcal{F}_\phi$ is very similar in $\nu=0.8\lambda$. The tendency of Fig. 6(b) and Fig. 6(d) is like in $\nu=4\lambda$.

\begin{figure}
	\includegraphics[width=5.5cm]{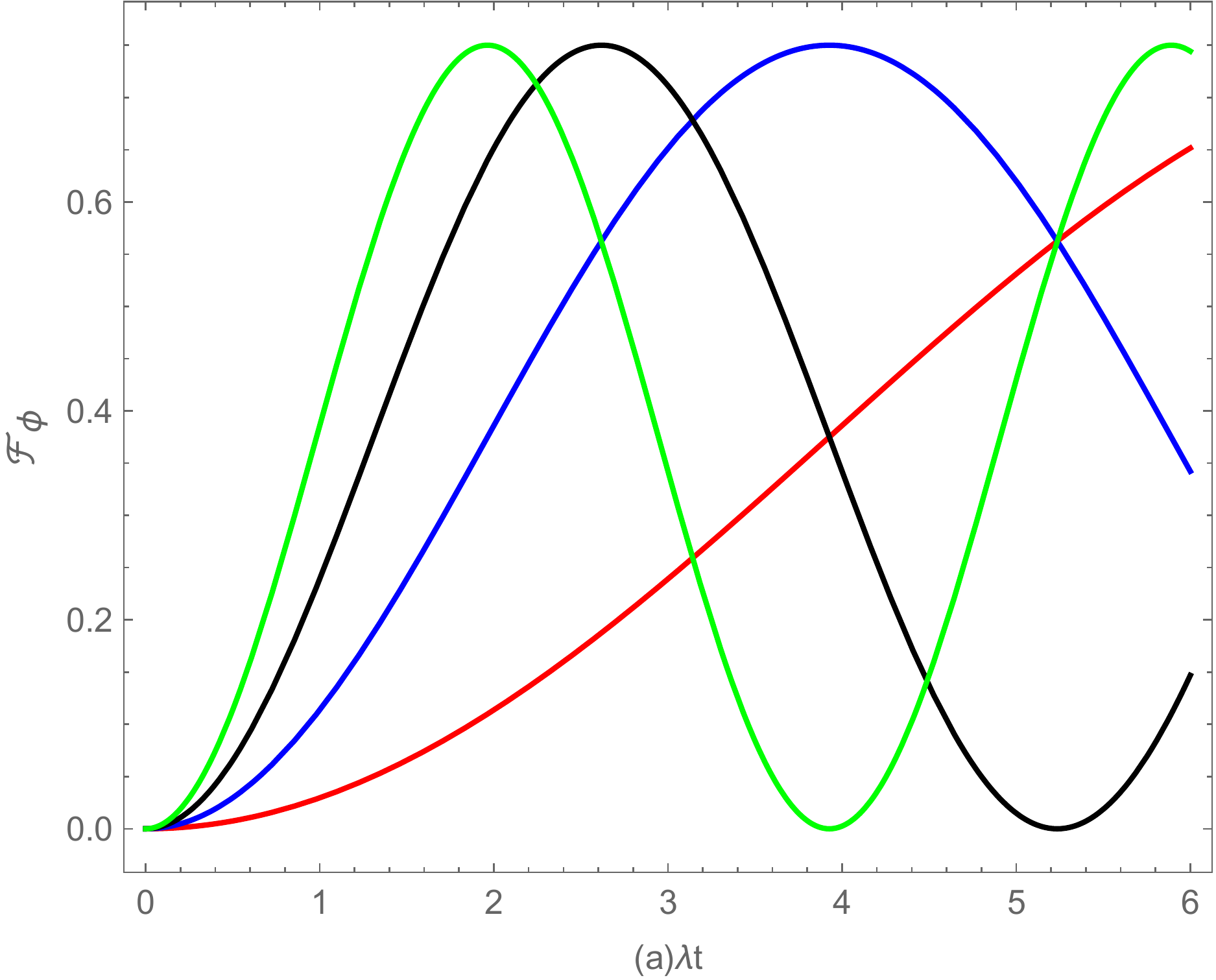}
	\includegraphics[width=5.5cm]{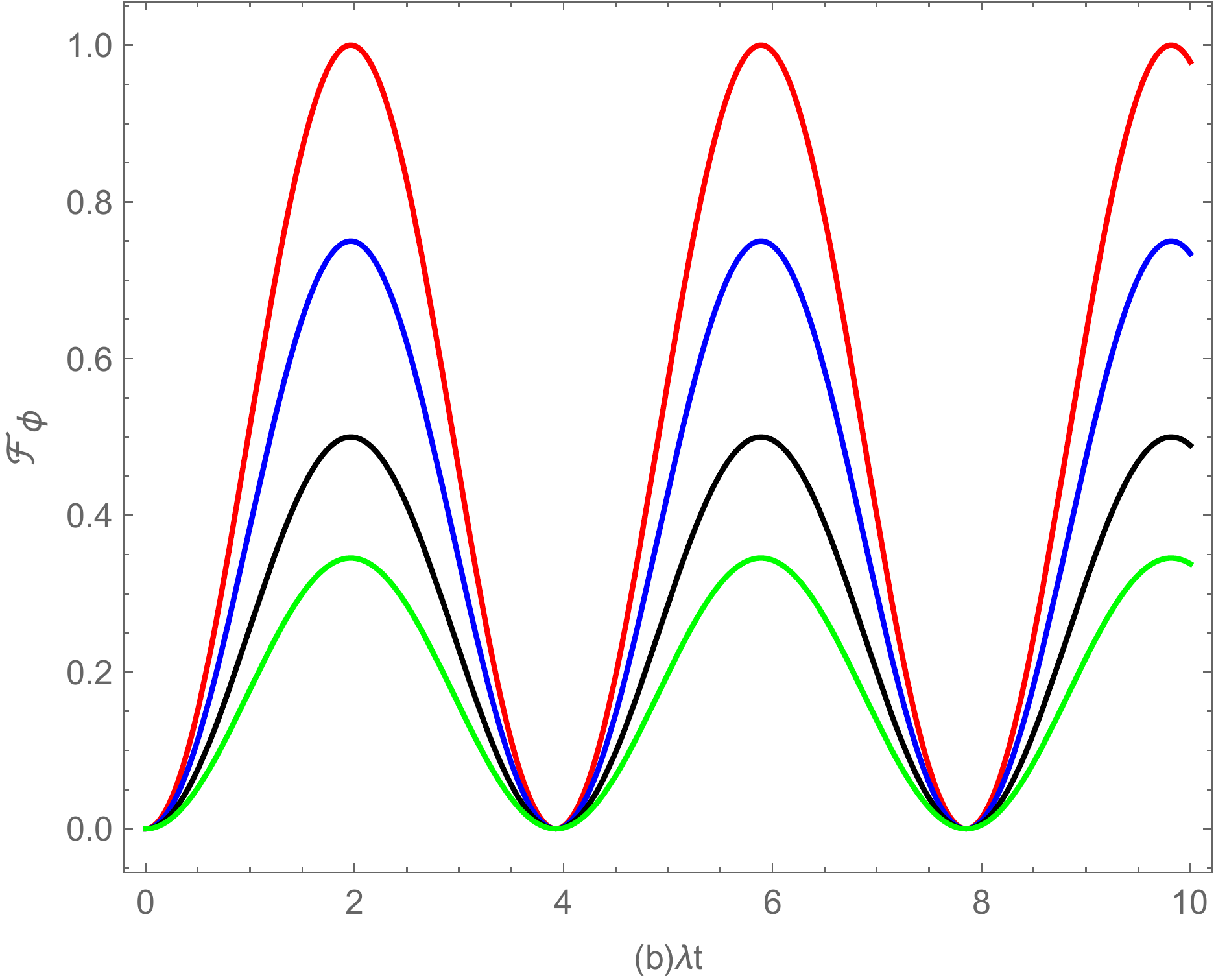}
	\caption{(Color online) Left: QFI $\mathcal{F}_\phi$ as a function of $\lambda t$ when $a=0$, $\kappa=0$, $\theta=\frac{\pi}{2}$ and under different values of $\nu$: $\nu=0.2\lambda$(red line), $\nu=0.4\lambda$(blue line), $\nu=0.6\lambda$(black line), $\nu=0.8\lambda$(green line). Right: QFI $\mathcal{F}_\phi$ as a function of $\lambda t$ when $a=0$, $\kappa=0$, $\nu=0.8\lambda$ and under different values of $\theta$: $\theta=\frac{\pi}{2}$(red line), $\theta=\frac{\pi}{3}$(blue line), $\theta=\frac{\pi}{4}$(black line), $\theta=\frac{\pi}{5}$(green line).}
	\label{fig:7}       
\end{figure}

In the above Figures 1-6, we have discussed in detail the influences of different noise parameters on the dynamical behaviors of QFI and QC of the qubit in a nonequilibrium environments. In order to better understand the difference between nonequilibrium and equilibrium environments, we also draw the dynamical curves of the QFI of the qubit in the equilibrium environment, as shown in the Figures 7(a) and (b). Figures 7(a) shows that these curves share the same beginning for different $\nu$ values and Figures 7(b) indicates that these curves also share the same beginning for different initial states when the nonequilibrium environment parameter($a=0$) and the memory decay rate($\kappa=0$). In the equilibrium environment, the dynamical curves of $\mathcal{C}_{l}(t)$ are similar to that of the $\mathcal{F}_\phi$ so we omit it in order to reduce the space.

Comparing Figure 7 and Figures 1-6, we can know that, in the equilibrium environment, the different noise parameters only alter their time evolution but not their initial values. However, in the nonequilibrium environment, the different noise parameters alter not only their time evolution but also their initial values.

\section{Conclusion}

In summary, we have investigated the non-Markovianity, QFI and quantum coherence of a single qubit coupled to a nonequilibrium environment and have obtained the expressions of QFI and quantum coherence as well as the relationship between QFI and quantum coherence. Their relationship ( Eq.~(\ref{EB20})) indicates that the quantum coherence can enlarge the QFI and can effectively enhance the quantum metrology, which gives us a novel understanding of the relationship among quantum resources. We have also discussed in detail the influences of the different noise parameters on the non-Markovianity, QFI and quantum coherence. The results show that, the suitable environment parameters, including the nonequilibrium parameter, the memory decay rate and the jumping rate, are beneficial to enhance the non-Markovianity and slower the decaying of QFI and quantum coherence. And we utilized the non-Markovianity to explain the dynamical behaviors of the QFI and the quantum coherence. Besides, in the equilibrium environment, the different noise parameters only alter their time evolution but not their initial values. However, in the nonequilibrium environment, the different noise parameters alter not only their time evolution but also their initial values.

These results give us an active way to suppress decoherence, which is quite significant in quantum information processing and quantum metrology. These would be a very interesting issue that needs further attention. We hope that our work is valuable to understand the common feature of quantum resources and the relationship between QFI and quantum coherence. In future work, we will study the essential relations between quantum resources and try to provide a unified framework of them.

\section*{Acknowledgments}
Project supported by the Scientific Research Project of Hunan Provincial Education Department, China (Grant No 16C0949) and the National Science Foundation of China (Grant No 11374096).

\end{document}